\documentstyle[aps,prl] {revtex}
\twocolumn
\begin{document}

\title{Six-Dimensional Quantum Dynamics of Adsorption and Desorption
       of H$_2$ at Pd(100):\\ Steering and Steric Effects}

\author{Axel Gross, Steffen Wilke, and Matthias Scheffler}

\address{Fritz-Haber-Institut der Max-Planck-Gesellschaft, Faradayweg 4-6,
D-14195 Berlin-Dahlem, Germany}

\maketitle

\begin{abstract}

We report the first six-dimensional quantum dynamical calculations of
dissociative adsorption and associative desorption. Using a potential
energy surface obtained by density functional theory calculations, we show
that the initial decrease of the sticking probability with increasing kinetic
energy
in the system H$_2$/Pd(100), which is usually attributed to the existence of
a molecular adsorption state, is due to dynamical steering.
In addition, we examine the influence of rotational motion and
orientation of the hydrogen molecule on adsorption and desorption.

\end{abstract}


The dissociative adsorption and associative desorption of hydrogen on
transition
metals is of direct technological relevance as well as of general
interest for the very fundamental understanding of elemental chemical
processes occurring in catalysis and surface chemistry.
Molecular beam experiments of the dissociative adsorption of H$_2$ on various
transition metal surfaces like Pd(100) \cite{Ren89}, Pd(111) and Pd(110)
\cite{Res94}, W(111) \cite{Ber92}, W(100) \cite{Ber92,But94,Aln89}, and
Pt(100) \cite{Dix94}  revealed
that the sticking probability initially decreases with increasing kinetic
energy
of the beam. Such a behavior is usually explained by a {\em precursor
mechanism}
\cite{Ren94}. In this concept the molecules are temporarily trapped in a
molecular physisorption state, the so-called {\em precursor state},
before they dissociatively adsorb.
The trapping occurs because the molecules lose energy when impinging on
the surface, mainly to the substrate phonons \cite{Ren94}.
The energy dependence of the sticking probability is thus related to the
trapping probability into the precursor state, and
it is this trapping probability which decreases with increasing energy.
In all experiments mentioned above the sticking probability rises again
at higher energies which is interpreted as such that
adsorption via direct activated paths then becomes dominant.

The relevance of the precursor mechanism for the dissociative adsorption
has been discussed for a long time, and the discussion remains
quite lively \cite{SSSS8,Lun94}. One may argue that especially for
hydrogen adsorption on metal surfaces the energy transfer to the phonon
system should be relatively small due to the large mismatch in the masses of
adsorbate and substrate making the precursor mechanism ineffective.
Therefore also direct non-activated adsorption together with a dynamical
steering effect has been suggested in order to explain the initial decrease
of the sticking probability (see, e.g., \cite{Aln89,Dix94}), however, there
had been no theoretical analysis of the importance of
this effect. Recent density functional theory calculations \cite{Wil95}
of the potential energy surface (PES) of H$_2$/Pd(100) show that there
exist non-activated as well as activated paths to dissociative adsorption in
this system, but they do not yield any {\em molecular}
adsorption well. These results motivated
dynamical calculations using a model PES with activated as well as
non-activated
path to dissociative adsorption \cite{Gro95}, in which
three degrees of freedom of the hydrogen molecule were taken into
account. This investigation suggested that the steering mechanism could
be operative in the system H$_2$/Pd(100), but the results showed
large quantitative discrepancies to the experiment \cite{Gro95}.

In order to clarify the nature of the dissociation process, we parametrized
the {\it ab initio} PES of H$_2$/Pd(100) thereby obtaining a well-defined
function describing the potential energy surface on which the hydrogen
molecule
moves. This function depends on six coordinates, and together with the
kinetic
energy operator it defines the Hamiltonian for the nuclear motion. The
Hamiltonian is then used for a quantum dynamical study
where for the first time {\em all} six degrees
of the hydrogen molecule have been treated fully quantum mechanically in a
simulation of dissociative adsorption and associative desorption.
The specific coordinates considered in our study
are the three coordinates of the center of mass of
the molecule $X$, $Y$, and $Z$ ($Z$ is the distance from the surface),
the azimuthal orientation $\phi$ and the polar orientation $\theta$,
and the H-H distance $d$. The potential in the $Zd$ plane is
described in reaction path coordinates $s$ along the reaction path and
$r$ perpendicular to it \cite{Bre89}. The potential is then given in the
following form
\begin{equation}
V \ (X,Y,s,r,\theta,\phi) \ = \ V^{corr} \ + \ V^{rot} \ + \ V^{vib}
\end{equation}
with
\begin{equation}
V^{corr} \ = \ \sum_{m,n = 0 }^2 \ V_{m,n}^{(1)} (s) \ \cos mGX \ \cos nGY,
\end{equation}
\begin{eqnarray}
\lefteqn{V^{rot} = \sum_{m=0}^1 \ V_m^{(2)}(s) \ \frac{1}{2}
\cos^2 \theta \  (\cos mGX + \cos mGY)} \nonumber\\
& & + \sum_{n=1}^2 \ V_n^{(3)} (s) \ \frac{1}{2}
 \sin^2 \theta \ \cos 2 \phi \ (\cos nGX - \cos nGY)
\end{eqnarray}
and
\begin{equation}
V^{vib} \ = \ \frac{\mu}{2} \ \omega^2 (s) \ [r \ - \ \Delta r (X,Y,s)]^2.
\end{equation}
$G = 2 \pi / a$ is the basis vector of the reciprocal lattice, $a$ is the
lattice constant, $\mu$ is the reduced mass of the hydrogen molecule,
and $\omega (s)$ is the frequency perpendicular to the reaction path.
The functions $V_{m,n}^{(i)}(s)$, $\omega (s)$, and $\Delta r (X,Y,s)$
are determined such that the difference between $V (X,Y,s,r,\theta,\phi)$
and the {\it ab initio} total energies, which have been calculated for more
than 250 configurations, on the average is smaller than 25~meV.

Although the PES does not have a precursor molecular adsorption state,
the results of the six-dimensional quantum dynamical calculations concerning
the dependence of the sticking probability on the kinetic energy
in the system H$_2$/Pd(100) agree very well with the measured results.
This will be explained by a purely dynamically steering effect.
In addition we will show that
rotations in general suppress sticking, but that steric effects,
i. e., effects due to the orientation of the molecules, can lead to a
strong enhancement of the sticking probability.

The quantum dynamics is determined by solving the time-independent
Schr\"odinger equation for the two hydrogen nuclei moving on the
six-dimensional PES.
We use the concept of the {\em local reflection matrix} (LORE) \cite{Bre93}
and the {\em inverse local transmission matrix} (INTRA) \cite{Bre94}.
This very stable coupled-channel method has been employed before
in a high-dimensional study of the adsorption of H$_2$/Cu(111), where,
however,
the polar orientation of the molecule has not been varied \cite{Gro94}.
In the present calculation up to 21,000 channels per total energy are taken
into
account; with channels the eigenfunctions of the asymptotic Hamiltonian
are meant. The PES is obtained
using density functional theory within the generalized gradient approximation
(GGA) \cite{Per92} and the full-potential linear augmented plane wave method
(FP-LAPW) (see Ref.~\cite{Wil95} and references therein).

Figure \ref{stick} presents our results for the sticking
probability as a function of the kinetic energy of the incident H$_2$ beam.
The dashed curve corresponds to H$_2$ molecules initially in the
rotational ground state $j_i = 0$.
This curve exhibits a strong oscillatory structure for low energies
in contrast to the experimental results. Such oscillatory
structures have already been found in three-dimensional model calculations
using potential energy surfaces with activated as well as non-activated paths
to adsorption \cite{Gro95,Dar90}. For a PES without a molecular adsorption
well the oscillations are related to quantum effects of the hydrogen beam,
namely
the opening up of new scattering channels for increasing incident energy
\cite{Gro95}.

In molecular beam experiments the incident molecules are not all in the
rotational ground state, the rotational states are occupied according to
a Boltzmann distribution with a temperature of 0.8 of the nozzle temperature
\cite{Ren89}. In addition, the beam is not strictly monoenergetic,
but has a certain energy spread. Both effects, the rotational distribution
and the energy spread, are incorporated in the solid curve where we
have assumed a typical energy spread of molecular beams of
$\Delta E / E_i = 2 \Delta v / v_i = 0.2$ \cite{Ren89} ($E_i$ and $v_i$ are
the initial kinetic energy and velocity, respectively). This curve should be
compared with the experimental results of Rendulic~{\it et al.} \cite{Ren89}.
In view of the fact that {\em no} parameters are used in our calculations,
the agreement between theory and experiment is quite satisfactory.
The initial decrease of the sticking probability is well reproduced although
no precursor state exists in our PES. Hence a purely
dynamical mechanism has to be the origin of this energy dependence.

The mechanism, a dynamical steering effect, is disclosed by the fact that
more channels in the dynamical simulation are needed at low energies
than at high energies. Usually it is the other way around in
coupled-channel calculations since the higher the energy, the more channels
become energetically accessible. Thus the calculations reflect that there is
a strong rearrangement between the different channels at low energies due to
the steering. At high energies the dynamics is closer to the adiabatic
limit which means that during the dissociation event there is little
redistribution among the channels.

An illustration of the steering effect is given in fig.~\ref{PES}.
There a two-dimensional cut through the six-dimensional PES is shown together
with two typical trajectories of impinging H$_2$ molecules.
Besides a difference in velocity, both trajectories have equivalent
impact parameters. In the case of the left trajectory the incident molecule
is so slow that the forces acting on the molecule can steer it towards a
configuration where non-activated dissociative adsorption is possible,
while in the case of the right trajectory the molecule is too fast for the
forces to divert the molecule significantly, it hits the repulsive part of
the potential and is reflected back into the gas phase.
If the incident energy of the molecule
is further increased, from a certain energy on it will have enough
energy to directly traverse the barrier. This leads to the increase of
the sticking probability at higher energies (see fig.~\ref{stick}).

Of course in a complete description of the adsorption process
also energy losses of the impinging molecules to the substrate
phonons have to be considered, in spite of the large mass mismatch
which makes the energy transfer rather small.
Energy losses of the molecules would slow down the molecule and thus
make the steering even more effective. Furthermore, the inclusion of
the substrate motion would broaden the velocity spread of the incident
molecules relative to the substrate atoms \cite{Han90}.

Figure~\ref{stick} also demonstrates that by taking into account the
rotational
population of the incoming beam in adsorption, the averaged sticking
probability is slightly decreased as compared to molecules in
the rotational ground state. This effect is shown in more detail in
fig.~\ref{steric}, which displays the sticking probability versus initial
rotational quantum state for a fixed incident kinetic energy of
$E_{i} = 0.19$~eV. The diamonds correspond to the orientationally averaged
sticking probability
\begin{equation}\label{ave}
\bar S_{j_i} (E) \ = \ \frac{1}{2j_i +1} \ \sum_{m_i = -j_i}^{j_i} \
S_{j_i,m_i} (E),
\end{equation}
which decreases with increasing $j_i$.
This means that the faster the molecules are rotating, the more the
dissociative adsorption is suppressed, and is caused
by {\em orientational hindering} \cite{Dar94}:
molecules with a high angular momentum will rotate out of a favorable
orientation towards adsorption during the dissociation event.

{}From fig.~\ref{steric} and the principle
of microscopic reversibility, it follows that the population of
rotational states in desorption should be lower than expected for
molecules in thermal equilibrium with the surface temperature.
This so-called rotational cooling has indeed been found for H$_2$
molecules desorbing from Pd(100) \cite{Sch91}.
Figure~\ref{desrot} shows a comparison between theory and experiment
of the rotational temperatures $T_{rot}$ of desorbing molecules.
These temperatures are determined by fitting the rotational state population
in a Boltzmann plot to a straight line. The slope then gives
$(-k_B T_{rot})$.
In order to compare the calculations with the experimental data
which were limited to low $j$ states, for the theoretical
analysis only the lowest four rotational states have been considered.
There is again a rather good agreement between theory and experiment.

The results of figs.~\ref{stick} and~\ref{desrot} have been determined by
summing over all azimuthal quantum numbers $m$. From the PES we know,
however, that the most favorable configuration towards dissociative
adsorption is met when the molecular axis is parallel to the surface.
Rotational eigenstates with azimuthal quantum number
$m =j$ have their rotational axis preferentially perpendicular to the
surface, so that the molecular axis is oriented mainly parallel to the
surface.  Molecules rotating in this so-called helicopter fashion should
dissociate more easily than molecules rotating in the cartwheel fashion
with the rotational axis preferentially parallel to the
surface ($m = 0$) since the latter have a high probability hitting the
surface in an upright orientation in which they cannot dissociate.
This steric effect has been investigated in a number of model studies
for activated adsorption employing three and four-dimensional quantum
dynamics \cite{Dar94,Mow93,Bru94,Dai95} or mixed classical-quantum
dynamics \cite{Kum94}.

In fig.~\ref{steric} also the sticking probability for initial rotational
quantum state for $m_i = 0$ and $m_i = j_i$ has been plotted.
The results show that the steric effect is operative in
systems with activated as well as non-activated paths towards dissociative
adsorption. Indeed the difference in the sticking probability
between initial $m_i = 0$ and $m_i =j_i$ rotations increases with increasing
quantum number $j_i$. This reflects the fact that the larger the rotational
quantum number $j$ is, the larger is the difference in the orientation
between
$m=0$ and $m=j$ states. The $m_i =j_i$ data demonstrate that the steric
effect even over-compensates the suppression of the sticking probability by
rotational motion since the $m_i = j_i$ sticking probability increases with
increasing $j_i$.

In conclusion, we reported the first six-dimensional quantum dynamical
study of dissociative adsorption on and associative desorption from surfaces.
We employed a potential energy surface obtained by detailed density
functional theory calculations for the system H$_2$/Pd(100).
The results show an initial decrease of the sticking
probability with increasing energy which, in contrast to common believe,
is here not due to a molecular adsorption state, but can be explained by
a dynamical steering effect. Rotational motion in general
suppresses sticking, however, a strong enhancement in the sticking
probability for molecules with their axis parallel to the surface
is found. The results clearly demonstrate that a high-dimensional
dynamical analysis of the PES is indeed important and provides,
due to the microscopic information, {\em quantitative} as well as
new {\em qualitative} understanding of processes at
surfaces.

\begin{figure}[h]

   \caption{Sticking probability versus kinetic energy for
a H$_2$ beam under normal incidence on a Pd(100) surface.
Dashed line: H$_2$ molecules initially in the rotational ground state;
Solid line: H$_2$ molecules with an initial rotational and energy
distribution adequate for molecular beam experiments (see text); circles:
experiment (from ref.~\protect{\cite{Ren89}}).}

\label{stick}
\end{figure}

\begin{figure}[h]
   \caption{Illustration of the steering effect. The contour
lines show the potential energy surface in eV along a two-dimensional cut
through the six-dimensional configuration space of a hydrogen molecule.
The molecular axis is parallel to the surface.
The Pd atoms are at the potential maxima.
The surface coordinate corresponds to the $X$-axis, while the reaction path
coordinate $s$ for $s \rightarrow \infty$ is equivalent to the H$_2$ center
of mass distance from the surface, for $s \le - 2.5$~{\AA} to the distance
between two adsorbed H~atoms.
The left trajectory corresponds to a slow molecule (kinetic energy
$E_{kin} = 0.05$~eV), the right trajectory to a fast molecule
($E_{kin} = 0.15$~eV).}

\label{PES}
\end{figure}

\begin{figure}[h]
   \caption{Sticking probability versus initial rotational quantum state
   $j_i$. Diamonds: orientationally averaged sticking probability
   (eq.~\protect{\ref{ave}}),
   triangles: $m_i = 0$ (cartwheel rotation),
   circles: $m_i = j_i$ (helicopter rotation).
   The initial kinetic energy is $E_{i} = 0.19$~eV.}

\label{steric}
\end{figure}

\begin{figure}[h]
   \caption{Rotational temperatures of desorbing H$_2$ molecules.
     Circles: experiment (from ref. \protect{\cite{Sch91}}), diamonds:
theory.
     The solid line corresponds to molecules in equilibrium with the
surface temperature.}

\label{desrot}
\end{figure}

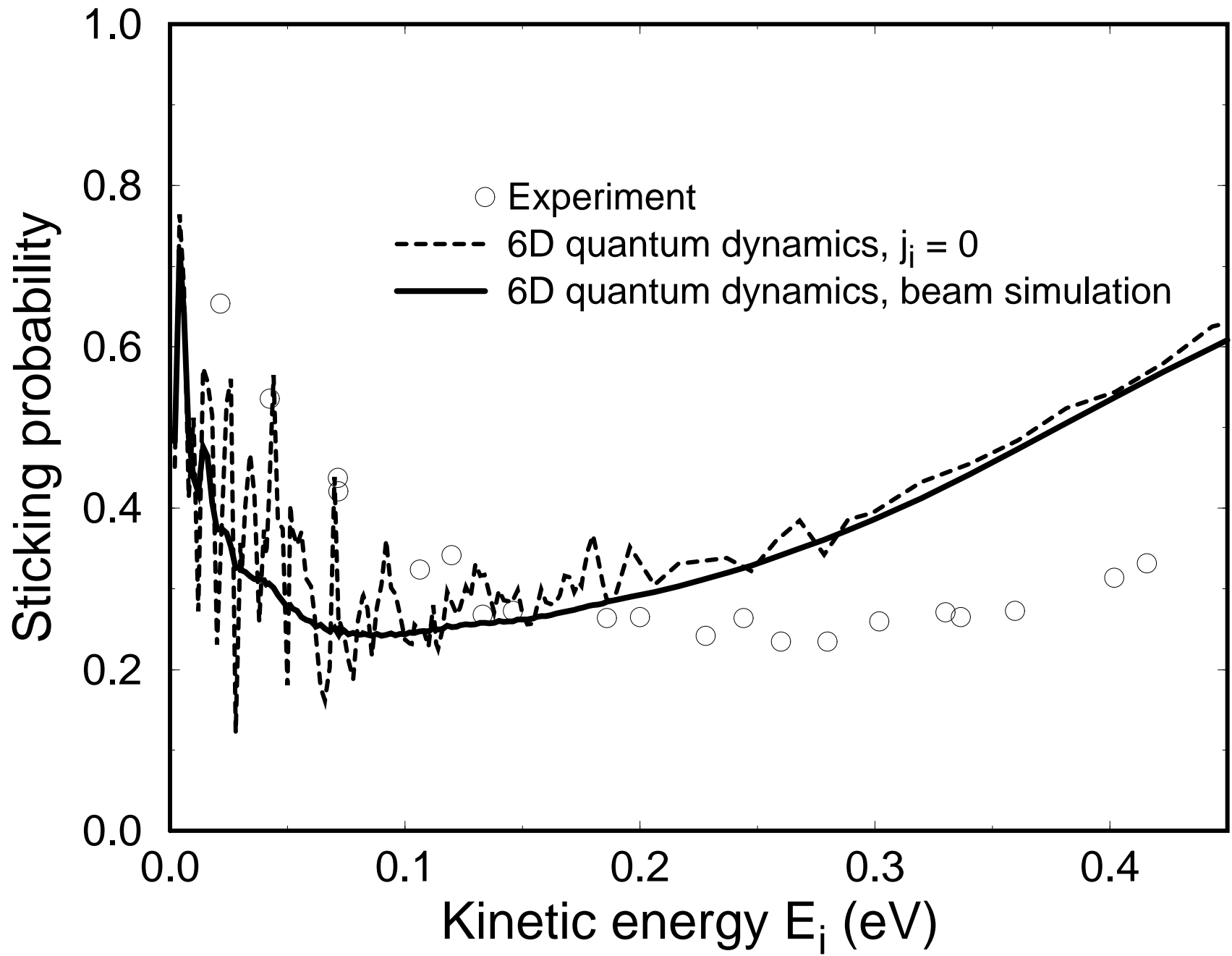
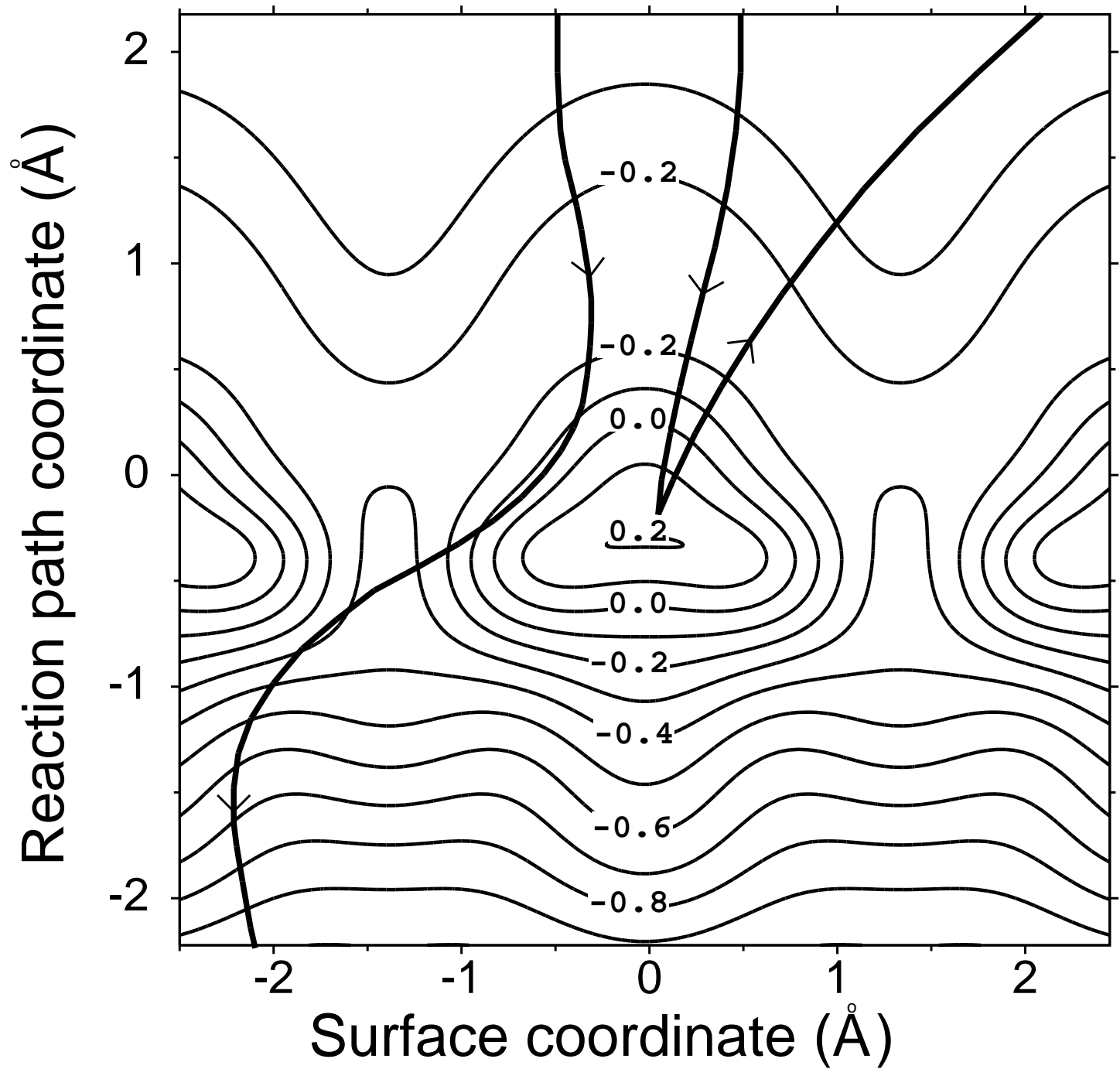
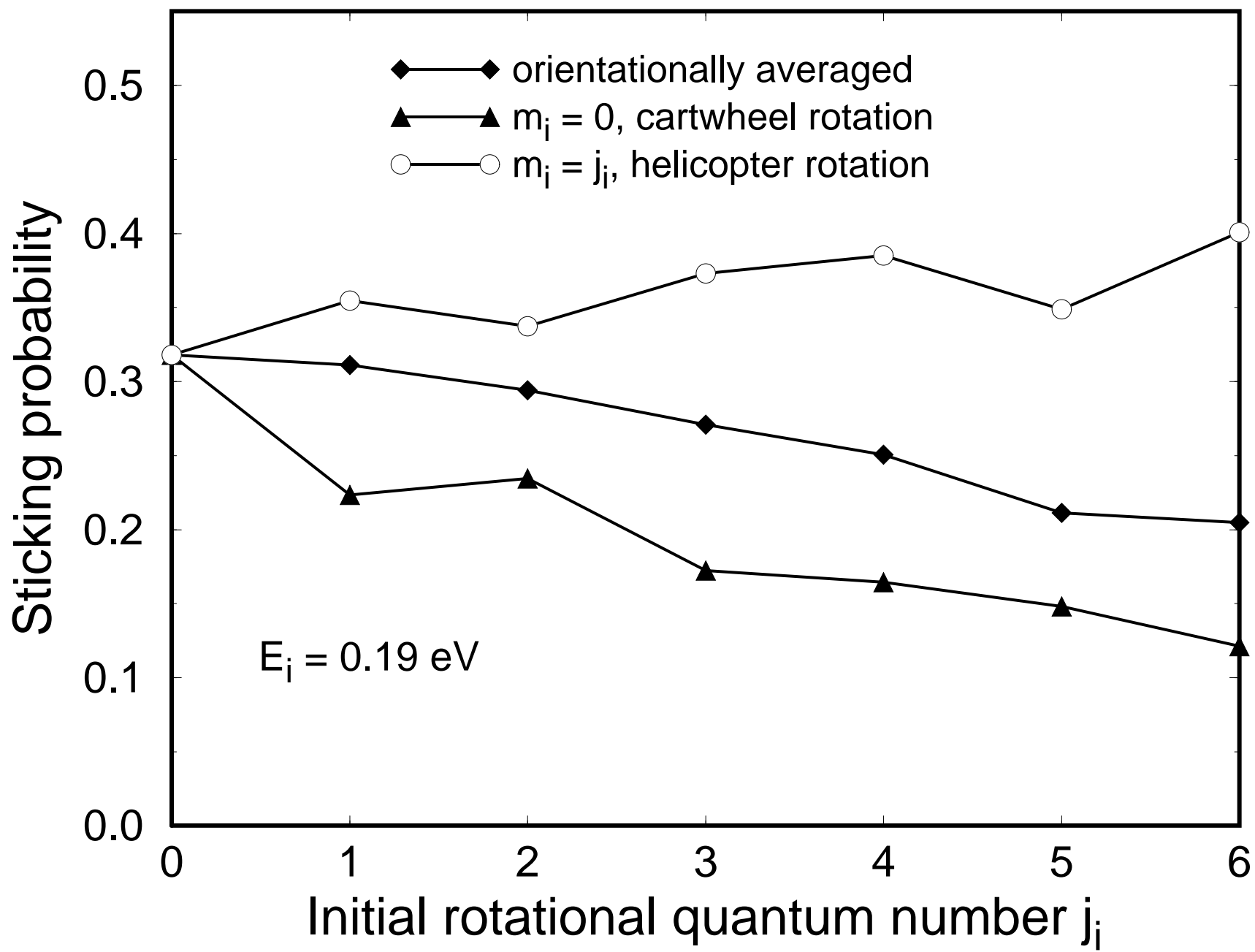
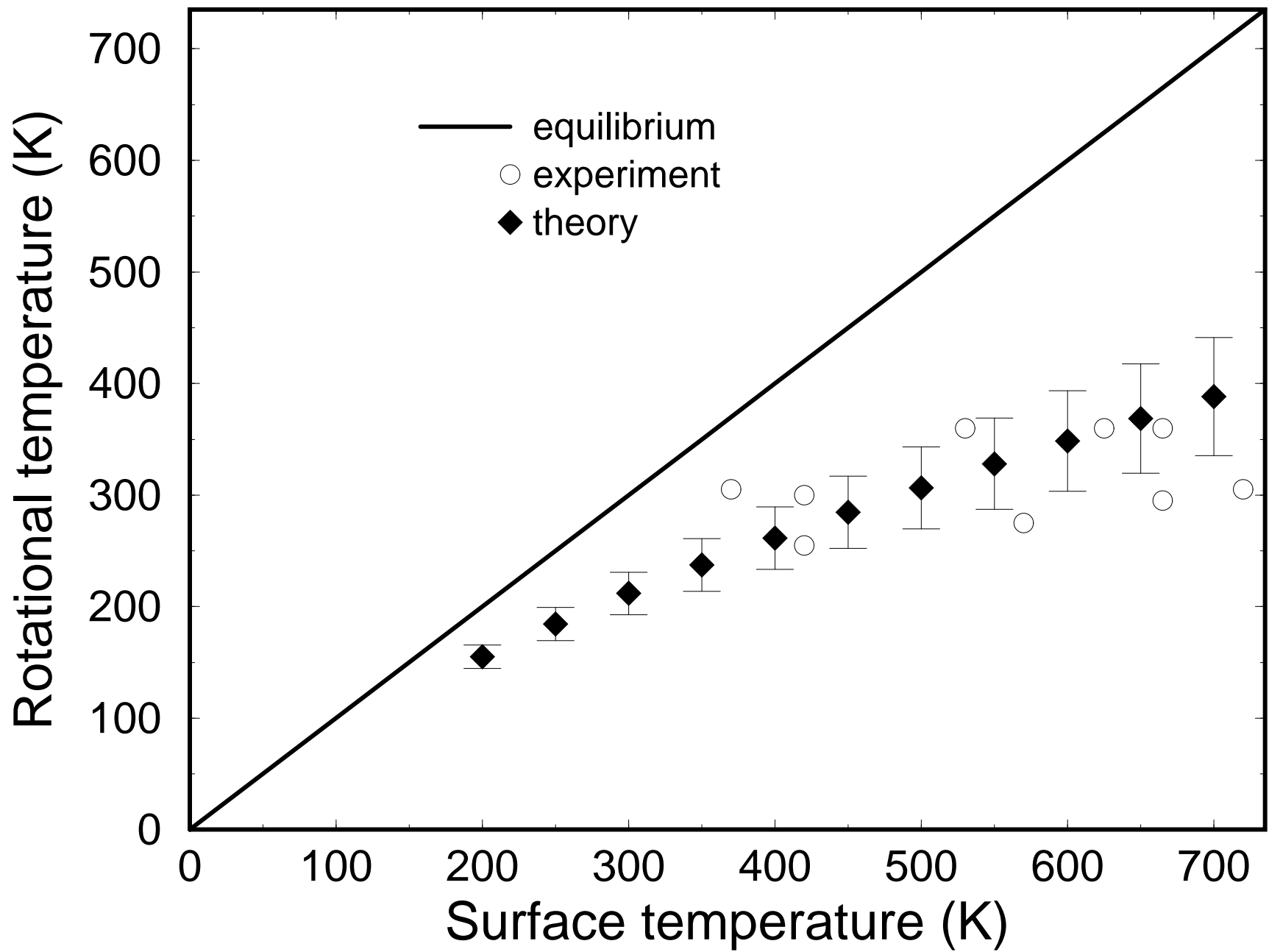
\end{document}